# NEW DEVELOPMENT OF REALISTIC *J*-MATRIX INVERSE SCATTERING *NN* INTERACTION AND *AB INITIO* DESCRIPTION OF LIGHT NUCLEI

**A. M. Shirokov[1,2], V. A. Kulikov[1], P. Maris[2], A. I. Mazur[3], E. A. Mazur[3], J. P. Vary[2]**

[1]*Skobeltsyn Institute of Nuclear Physics, Moscow State University, Moscow, Russia*
[2]*Iowa State University, Ames, IA, USA*
[3]*Pacific National University, Khabarovsk, Russia*

Abstract: We discuss the studies of light nuclei in *ab initio* No-core Full Configuration approach based on extrapolations to the infinite model space of large-scale No-core Shell Model calculations on supercomputers. The convergence at the end of *p* shell and beginning of the *sd* shell can be achieved if only reasonable soft enough *NN* interactions are used. In particular, good predictions are obtained with a realistic JISP16 *NN* interaction obtained in *J*-matrix inverse scattering approach and fitted to reproduce light nuclei observables without three-nucleon forces. We discuss the current status of this *NN* interaction and its recent development.

## 1. Introduction

A rapid development of *ab initio* methods for solving finite nuclei has opened a range of nuclear phenomena that can be evaluated to high precision using realistic nucleon-nucleon interactions. Nowadays, due to increased computing power and novel techniques, *ab initio* approaches like the No-core Shell Model (NCSM) [1], the Green's function Monte Carlo [2] and the Coupled-Cluster Theory [3] are able to reproduce properties of a large number of atomic nuclei with mass up to $A$=16 and can be extended for heavier nuclei. Recently a new *ab initio* method, the No-Core Full Configuration (NCFC) approach [4], was introduced. NCFC is based on extrapolation of the NCSM results in successive basis spaces to the infinite basis space limit. This makes it possible to obtain basis space independent predictions for binding energies and to evaluate their numerical uncertainties. We concentrate the discussion here on the NCFC approach and on some new results obtained with it. In particular, we discuss the predictions for the binding energy and spectrum of the extreme proton-excess nucleus $^{14}$F [5] for which the first experimental observation is expected to be reported soon.

The *ab initio* methods require a reliable realistic strong interaction providing an accurate description of *NN* scattering data and high-quality predictions for binding energies, spectra and other observables in light nuclei. A number of meson-exchange potentials sometimes supplemented with phenomenological terms to achieve high accuracy in fitting *NN* data (CD-Bonn [6], Nijmegen [7], Argonne [8]) have been developed that should be used together with modern *NNN* forces (Urbana [9,10], Illinois [11], Tucson–Melbourne [12,13]) to reproduce properties of many-body nuclear systems. On the other hand, one sees the emergence of *NN* and *NNN* interactions with ties to QCD [14–17].

Three-nucleon forces require a significant increase of computational resources needed to diagonalize a many-body Hamiltonian matrix since the *NNN* interaction increases the number of non-zero matrix elements approximately by a factor of 30 in the case of *p*-shell nuclei. As a result, one needs to restrict the basis space in many-body calculations when *NNN* forces are involved that makes the predictions less reliable. *Ab initio* many-body studies benefit from the use of recently developed purely two-nucleon interactions of INOY (Inside Nonlocal Outside Yukawa) [18,19] and JISP (*J*-matrix Inverse Scattering Potential) [20–23] types fitted not only to the *NN* data but also to binding energies of $A$=3 and heavier nuclei. At the fundamental level, these *NN* interactions are supported by the work of Polyzou and Glöckle who demonstrated [24] that a realistic *NN* interaction is equivalent at the $A$=3 level to some *NN* + *NNN* interaction where the new *NN* force is related to the initial one through a phase-equivalent transformation (PET). It seems reasonable then to exploit this freedom and work to minimize the need for the explicit introduction of three- and higher-body forces. Endeavors along these lines have resulted in the design of INOY and JISP strong interaction models.

The JISP *NN* interaction provides a fast convergence of NCSM calculations, it is fitted in NCSM and NCFC studies to the properties of light nuclei and is developing together with the progress in these *ab initio* approaches. We discuss here the progress in developing of the JISP *NN* interaction in line with related progress of NCSM and NCFC studies of light nuclei.

## 2. JISP16 *NN* interaction and NCFC approach

The *J*-matrix inverse scattering approach was suggested in Ref. [25]. It was further developed and used to design a high-quality JISP *NN* interaction in Ref. [20]. A nonlocal interaction obtained in this approach is in the form of a matrix in the oscillator basis in each of the *NN* partial waves. To reproduce scattering data in a wider energy range, one needs to increase the size of the potential matrix and/or the $\hbar\Omega$ parameter of the oscillator basis. From the point of view of shell model applications, it is desirable however to reduce the size of potential matrices and to use $\hbar\Omega$ values in the range of few tens of MeV. A compromise solution is to use $\hbar\Omega$ = 40 MeV with $N_{\max}$ = 9 truncation of potential matrices [20], i. e., the JISP *NN* interaction matrices include all relative *NN* motion oscillator states with excitation quanta up to 8 or 9 depending on parity. In other words, we use potential matrices of the rank $r$ = 5 in *s* and *p* *NN* partial waves, $r = 4$ matrices in $d$ and $f$ partial waves, etc.; in the case of coupled waves, the rank of the potential matrix is a sum of the respective ranks, e. g., the rank of the coupled $sd$ wave matrix is $r = 5 + 4 = 9$. The $N_{\max} = 9$ truncated JISP interaction with $\hbar\Omega = 40$ MeV provides an excellent description of $NN$ scattering data with $\chi^2$/datum = 1.03 for the 1992 $np$ data base (2514 data), and 1.05 for the 1999 $np$ data base (3058 data) [26].

PETs originating from unitary transformations of the oscillator basis proposed in Refs. [27,28], give rise to ambiguities of the interaction obtained in the $J$-matrix inverse scattering approach. These ambiguities are eliminated at the first stage by postulating the simplest tridiagonal form of the $NN$ interaction in uncoupled and quasi-tridiagonal form in coupled $NN$ partial waves [20]. At the next stage, PETs are used to fit the JISP interaction to various nuclear properties. First of all, the $sd$ component of the $NN$ interaction is modified with the help of PETs to reproduce the deuteron quadrupole moment $Q$ and rms radius without violating the excellent description of scattering data. It is worth noting here that the deuteron binding energy $E_d$ and asymptotic normalization constants are used as an input in the inverse scattering approach and are not affected by PETs.

After that we employ PETs in other $NN$ partial waves attempting to improve the description of binding energies and spectra of light nuclei in NCSM calculations. Following this *ab exitu* route, the JISP6 $NN$ interaction fitted to properties of nuclei with masses $A \le 6$, was proposed [21]. It was found out later that JISP6 strongly overbinds nuclei with $A \le 10$. Therefore a new fit of PET parameters was performed that resulted in the JISP16 interaction [22] fitted to nuclei with masses up through $A \le 16$.

The JISP16 $NN$ interaction provides one of the best if not the best description of binding energies, spectra and other properties of $s$ and $p$ shell nuclei [22,4] as compared to other modern models of the realistic strong interaction. It is worth noting that JISP16 provides better convergence of *ab initio* calculations than other realistic $NN$ interactions and avoids the need to use three-nucleon forces. As a result, the JISP16 predictions for light nuclei are more reliable than that of other realistic models of $NN$ interactions. With modern supercomputer facilities, we can obtain converged or nearly converged energies of nuclei with mass $A \le 6$. For calculations of heavier nuclear systems, we proposed recently a NCFC approach [4].

It was found [4] that binding energies of many light nuclei represent an exponential convergence pattern in $N_{\max}$, the excitation oscillator quanta characterizing the basis space of the NCSM. Therefore, we fit the set of ground state energies obtained with each fixed $\hbar\Omega$ value using the relation

$$E_{gs}(N_{\max}) = a\exp(-cN_{\max}) + E_{gs}(\infty), \tag{1}$$

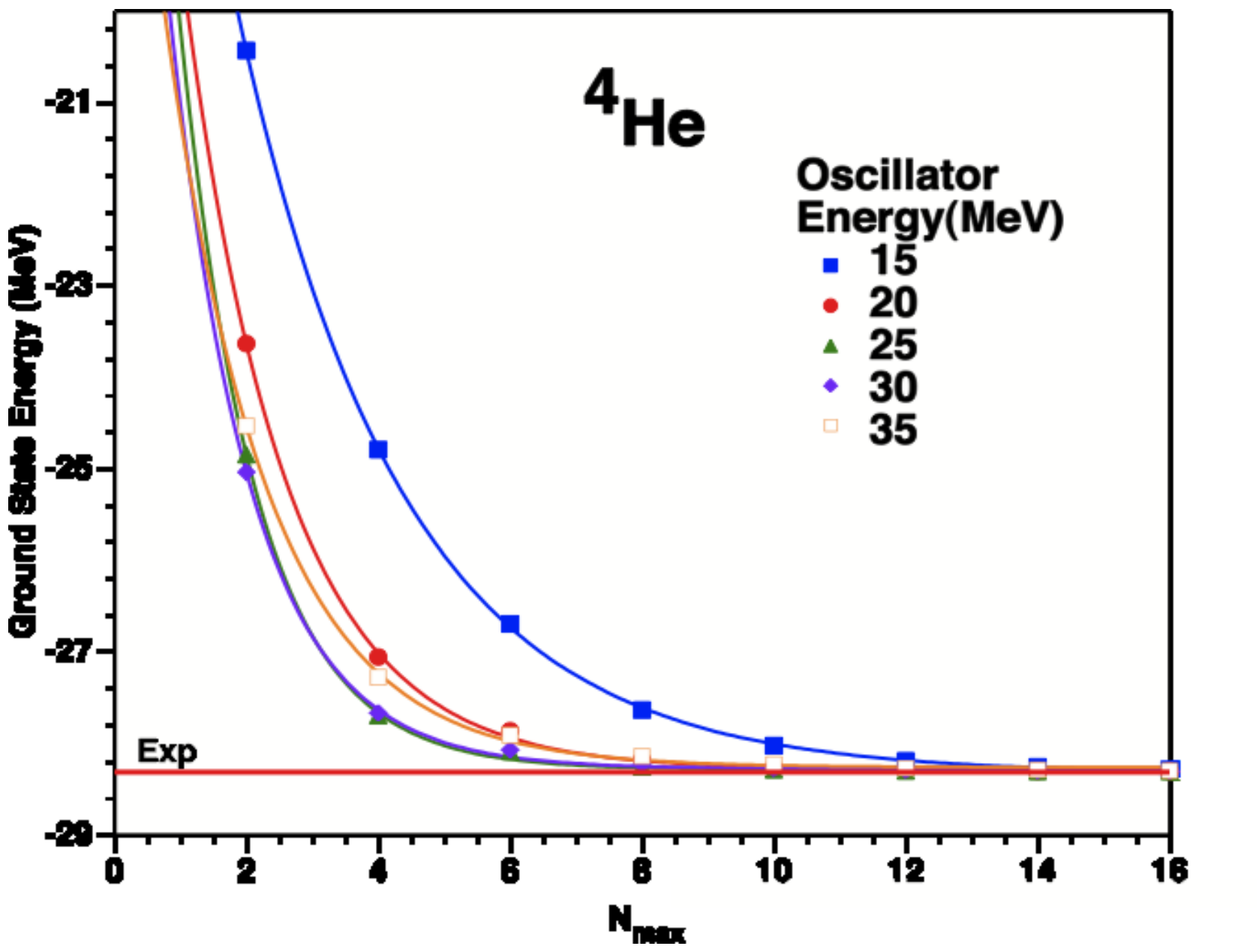

Figure 1. Ground state energies of $^4$He obtained with different $N_{\max}$ and $\hbar\Omega$ values. Each set of points at fixed $\hbar\Omega$ is fitted by Eq. (1) (solid curves).

where constants $a$ and $c$ depend on the $\hbar\Omega$ value and $E_{gs}(\infty)$ is the extrapolated ground state energy in the infinite basis space. The exponential convergence patterns and fits by Eq. (1) are illustrated by Fig. 1. Within the NCFC approach, we use two extrapolation methods: a global extrapolation based on the results obtained in four successive basis spaces with five $\hbar\Omega$ values from a 10 MeV interval (extrapolation A); and extrapolation B based on the results obtained at various fixed $\hbar\Omega$ values in three successive basis spaces and defining the most reliable $\hbar\Omega$ value for the extrapolation. These extrapolations provide consistent results and were carefully tested in a number of light nuclei where a complete convergence can be achieved [4].

An exciting recent result obtained with JISP16 *NN* interaction and NCFC method, is an *ab initio* prediction of properties of the exotic extreme proton-excess nucleus $^{14}$F. The first experimental results regarding this isotope will be available soon from Cyclotron Institute at Texas A&M University [29]. The largest calculations were performed in the $N_{\max}\hbar\Omega$ basis space with $N_{\max} = 8,$ which for this nucleus contains 1,990,061,078 basis states with natural parity (negative). The determination of the lowest ten to fifteen eigenstates of the sparse Hamiltonian matrix, for each oscillator parameter $\hbar\Omega,$ requires 2 to 3 hours on 7,626 quad-core compute nodes at the Jaguar supercomputer at ORNL.

We present in Table 1 the results of NCFC calculations [5] of the $^{14}$F ground state energy. Combining the extrapolations A and B predictions suggests a binding energy of $72 \pm 4$ MeV for $^{14}$F. To check the accuracy of our approach, we performed similar calculations for the mirror nucleus $^{14}$B with a known binding energy of 85.423 MeV [30]. This value agrees with our prediction of $86 \pm 4$ MeV. We also performed NCFC calculations of the neighboring nucleus $^{13}$O using basis spaces up to $N_{\max} = 8$. The calculated binding energy of $77 \pm 3$ MeV also agrees with the experimental value of 75.556 MeV [30].

Table 1. NCFC predictions for the ground state energies (in MeV) of $^{13}$O, $^{14}$B and $^{14}$F based on NCSM calculations with JISP16 in up to $N_{\max} = 8$ basis spaces. Estimates of the accuracy of the extrapolations are shown in parentheses. Experimental data are taken from Ref. [30].

| Nucleus | Extrapolation A | Extrapolation B | Experiment |
|---|---|---|---|
| $^{13}$O | –75.7(2.2) | –77.6(3.0) | –75.556 |
| $^{14}$B | –84.4(3.2) | –86.6(3.8) | –85.423 |
| $^{14}$F | –70.9(3.6) | –73.1(3.7) | ? |

We note that a good description of both $^{14}$F and $^{13}$O in the same approach is important in order to have a description of $^{14}$F consistent with the experiment in which $^{14}$F will be produced in the $^{13}\mathrm{O} + p$ reaction. In this way, any experimentally observed resonances can be directly compared with the difference of our predictions for the $^{14}$F and $^{13}$O energies. In this respect it is interesting to note that although the total energies of the extrapolations A and B differ by about 2 MeV, the differences between the ground state energies of these three nuclei are almost independent of the extrapolation method. (The numerical uncertainty in these differences is unclear, but expected to be significantly smaller than the uncertainty in the total energies.)

We also performed calculations [5] of the $^{14}$F excitation spectrum in anticipation of the experimental results. We performed independent separate extrapolation fits for total energies of all states. The differences between the extrapolated total energies and the ground state energy is our prediction for the excitation energies. This approach was carefully tested in Ref. [5] in calculations of the $^{6}$Li spectrum where a good convergence can be achieved. We summarize our results for the excited states of $^{14}$F and $^{14}$B in Table 2. The spectra are rather dense and the spacing between energy levels is smaller than the quoted numerical uncertainty, which is that of the extrapolated total energies of the excited states. However, as discussed above, we expect that for narrow resonances the actual numerical error in the excitation energy is (significantly) smaller than the error in the total energy. We expect the five lowest excited states and probably the $(5^-,2)$ state to have smaller widths than other states quoted in Table 2 (see Ref. [5] for a detailed discussion).

Some of the excited states in $^{14}$B were observed experimentally. Unfortunately, the spin of most of these states is doubtful or unknown. Overall, our predicted excitation energies appear to be too large when compared with the experimental data; in particular our prediction for the excited $2^-$ state, the only excited state with a firm spin assignment, is about 1.5 MeV above the experimental value. However, the spin of the lowest five states agrees with experiment, except for the $2^-$ and $4^-$ being interchanged, assuming that the tentative experimental spin assignments are correct. It would also be very interesting to compare our predictions for the $^{14}$F binding energy and spectrum with the experimental data that are anticipated soon.

Table 2. NCFC predictions for the $^{14}$F and $^{14}$B excitation energies (in MeV) based on NCSM calculations with JISP16 in up to $N_{\max} = 8$ basis spaces. Estimates of accuracies of extrapolations or experimental data (Ref. [30]) are given in parentheses.

| State | *Ab initio* NCFC calculations with JISP16 | | | | Experiment | |
|---|---|---|---|---|---|---|
| | $^{14}$F | | $^{14}$B | | $^{14}$B | |
| $E_x(J^\pi, T)$ | Extrap. A | Extrap. B | Extrap. A | Extrap. B | $J^\pi$ | $E_x$ |
| $E_x(1^-,2)_1$ | 0.9(3.9) | 1.3(2.5) | 1.1(3.5) | 1.4(2.8) | $(1^-)$ | 0.74(4) |
| $E_x(3^-,2)_1$ | 1.9(3.3) | 1.5(4.6) | 1.7(2.9) | 1.4(4.6) | $(3^-)$ | 1.38(3) |
| $E_x(2^-,2)_2$ | 3.2(3.5) | 3.3(3.5) | 3.3(3.1) | 3.3(3.8) | $2^-$ | 1.86(7) |
| $E_x(4^-,2)_1$ | 3.2(3.2) | 2.8(4.8) | 3.1(2.9) | 2.7(4.8) | $(4^-)$ | 2.08(5) |
| | | | | | ? | 2.32(4) |
| | | | | | ? | 2.97(4) |
| $E_x(1^-,2)_2$ | 5.9(3.5) | 5.4(4.6) | 5.9(3.1) | 5.5(4.8) | | |
| $E_x(0^-,2)$ | 5.1(5.4) | 5.8(1.0) | 5.5(4.8) | 6.1(1.4) | | |
| $E_x(1^-,2)_3$ | 6.2(4.8) | 6.3(2.8) | 6.4(4.3) | 6.4(3.1) | | |
| $E_x(2^-,2)_3$ | 6.4(4.6) | 6.3(3.4) | 6.9(4.1) | 6.7(3.6) | | |
| $E_x(3^-,2)_2$ | 6.9(4.2) | 6.4(4.6) | 7.0(3.7) | 6.5(4.7) | | |
| $E_x(5^-,2)$ | 8.9(3.5) | 7.9(5.9) | 8.8(3.1) | 7.8(5.9) | | |

### 3. Refined JISP16$_{2010}$ interaction

The new *ab initio* NCFC approach provides much more reliable predictions for bindings. The NCFC extrapolation technique revealed some drawbacks of the JISP16 *NN* interaction that was fitted to nuclear observables before this technique was developed. In particular, it was found that the JISP16 interaction overbinds essentially nuclei with mass $A \geq 14$ and $N \approx Z$.

These deficiencies of the *NN* interaction can be addressed by a new fit of the PET parameters defining JISP interaction in the NCFC calculations. We refer to as JISP16$_{2010}$ the revised *NN* interaction obtained in this fit. The JISP16 and JISP16$_{2010}$ describe *NN* scattering data with the same accuracy; the same PET is used to define both these interactions in the *sd* partial wave, hence they predict the same deuteron properties. However PET parameters in other *NN* partial waves differ between JISP16$_{2010}$ and JISP16. We note also that JISP16 was defined only in the *NN* partial waves with momenta $J \leq 4$ while the JISP16$_{2010}$ is extended to all $J \leq 8$.

Table 3. Binding energies (in MeV) of some nuclei obtained with JISP16 and $\mathrm{JISP16}_{2010}$ *NN* interactions; the $N_{\max}$ columns show the largest NCSM basis space used for the extrapolations.

| Nucleus | Experim. | JISP16 | | | $\mathrm{JISP16}_{2010}$ | | |
|---|---|---|---|---|---|---|---|
| | | Extrap. A | Extrap. B | $N_{\max}$ | Extrap. A | Extrap. B | $N_{\max}$ |
| $^{3}$H | 8.482 | $8.369 \pm 0.001$ | $8.3695 \pm 0.0025$ | 18 | $8.369 \pm 0.010$ | $8.367^{+0.012}_{-0.007}$ | 14 |
| $^{3}$He | 7.718 | $7.665 \pm 0.001$ | $7.668 \pm 0.005$ | 18 | $7.664 \pm 0.011$ | $7.663 \pm 0.008$ | 14 |
| $^{4}$He | 28.296 | $28.299 \pm 0.001$ | $28.299 \pm 0.001$ | 18 | $28.294 \pm 0.002$ | $28.294^{+0.002}_{-0.001}$ | 14 |
| $^{8}$He | 31.408 | $29.69 \pm 0.69$ | $29.29 \pm 0.96$ | 10 | $30.30 \pm 0.46$ | $29.99^{+1.31}_{-1.06}$ | 10 |
| $^{6}$Li | 31.995 | $31.47 \pm 0.09$ | $31.48 \pm 0.03$ | 16 | $31.33 \pm 0.16$ | $31.34 \pm 0.07$ | 14 |
| $^{10}$B | 64.751 | $63.1 \pm 1.2$ | $63.7 \pm 1.1$ | 8 | $62.6 \pm 1.4$ | $63.4 \pm 1.5$ | 8 |
| $^{12}$C | 92.162 | $93.9 \pm 1.1$ | $95.1 \pm 2.7$ | 8 | $91.1 \pm 1.3$ | $92.3 \pm 2.9$ | 8 |
| $^{14}$C | 105.284 | $112.1 \pm 2.1$ | $114.3 \pm 6.0$ | 8 | $102.5 \pm 1.6$ | $104.8 \pm 3.6$ | 8 |
| $^{14}$N | 104.659 | $114.2 \pm 1.9$ | $115.8 \pm 5.5$ | 8 | $102.7 \pm 1.5$ | $104.7 \pm 3.1$ | 8 |
| $^{16}$O | 127.619 | $143.5 \pm 1.0$ | $150 \pm 14$ | 8 | $126.7 \pm 3.1$ | $129.6 \pm 6.1$ | 8 |

We compare binding energies obtained with JISP16 and $\mathrm{JISP16}_{2010}$ interactions in Table 3. It is seen that the new interaction essentially improves the description of the *p* shell nuclei. In particular, $\mathrm{JISP16}_{2010}$ provides nearly exact binding energies of nuclei with $10 \le A \le 16$ and only slightly underbinds some of lighter nuclei listed in Table 3.

We plan to explore the properties of the refined realistic nonlocal *NN* interaction $\mathrm{JISP16}_{2010}$ in systematic large-scale calculations of other light nuclei including the ones with $A > 16$ and away from $N \sim Z$, and to carefully study its predictions not only for the binding energies but also for the spectra, electromagnetic transitions and other observables. Our plan is also to tune the interaction to the description of phenomenological nuclear matter properties.

This work was supported by the US DOE Grants DE-FC02-09ER41582 and DE-FG02-87ER40371 and Russian Federal Agency of Education Contract P521. Computational resources were provided by DOE through the National Energy Research Supercomputer Center (NERSC) and through an INCITE award (David Dean, PI).